# GPU accelerated manifold correction method for spinning compact binaries

Chong-xi Ran, Song Liu and Shuang-ying Zhong*

School of Sciences, Nanchang University, Nanchang Jiangxi 330031, China.

*Corresponding author. E-mail: Zhongshuangying@ncu.edu.cn

**Abstract**

The conservative Post-Newtonian (PN) Hamiltonian formulation of spinning compact binaries has six integrals of motion including the total energy, the total angular momentum and the constant unit lengths of spins. The manifold correction method can effectively eliminate the integration errors accumulation in a long time. In this paper, the accelerated manifold correction method based on graphics processing unit (GPU) is designed to simulate the dynamic evolution of spinning compact binaries. The feasibility and the efficiency of parallel computation on GPU for spinning compact binaries have been confirmed by various numerical experiments. The numerical comparisons show that the accuracy on GPU execution of manifold corrections method has a good agreement with the execution of codes on merely central processing unit (CPU-based) method. The acceleration ability when the codes are implemented on GPU can increase enormously through the use of shared memory and register optimization techniques without additional hardware costs, implying that the speedup is nearly 13 times as compared with the codes executed on CPU for phase space scan (including $314\times314$ orbits). In addition, GPU-accelerated manifold correction method is used to numerically study how dynamics are affected by the spin-induced quadrupole-monopole interaction for black hole binary system.

## 1. Introduction

Spinning compact binaries consisting of neutron stars or black holes, as a high-nonlinear and non-integrable relativistic two-body problem, are the most promising sources for the gravitational wave detectors. The match data analysis of gravitational waves relies on the theory template in addition to the observer's azimuth (Kidder 1995; Buonanno et al. 2006). The highly eccentric orbits tend to be chaotic due to the large velocity in the post-Newton (PN) Hamiltonian of spinning compact binaries (Hartl and Buonanno 2005). The chaos phenomena of spinning compact binaries have a significant effect on gravitational waves detection, attracting broad attention (Levin 2000; Huang et al. 2014; Huang et al.

2016).

There are three disputes about the regular and chaotic dynamics of PN approximations spinning compact binaries. More than a decade ago, Levin (2000) confirmed that the 2PN Lagrangian dynamics of a comparable-mass binary system with one spinning body (or two spins restricted to the leading order spin-orbit interaction) is chaotic by using the method of fractal basin boundaries. However, Schnittman and Rasio (2001) did not seek out chaos by finding no the positive Lyapunov exponents in recalculating Levin’s work. Shortly thereafter, Cornish and Levin (2003) obtained positive Lyapunov exponents using the method of two nearby trajectories. Why acquire the above contrary results for the same research problem? The primary cause of the contrary results is that they applied different chaotic indicators for seeking chaos, respectively. This was clarified by Wu and Xie (2007). They used the invariant Lyapunov exponents with the method of two nearby trajectories (Wu and Huang 2003) and the invariant fast Lyapunov indicators with the method of two nearby trajectories (Wu et al. 2006) to show that the different chaotic indicators lead to different dynamical behaviour. The second dispute is the relationship between the presence of chaos and the dynamic parameters or initial conditions of binary system. Due to no common rule for determining the presence of chaos, some works (Levin 2003; Hartl and Buonanno 2005) presented results in conflict with each other. For example, Levin (2003) declared that the binary system transforms from order to chaos with the increase of the spin magnitudes and misalignments, while the large eccentricity cannot cause chaos by itself, and the chaotic intensity is enhanced when the spins are perpendicular to the orbital angular momentum. Nevertheless, Hartl and Buonanno (2005) confirmed that the presence of chaos depends mainly on the initial spin vectors nearly anti-aligned with the orbital angular momentum in the equal-mass binaries, and chaos occurs in highly eccentric orbits. In fact, their results are not inconsistent, as stated in the references (Wu and Xie 2008). The third dispute is whether the 2PN Lagrangian and Hamiltonian approximations spinning black hole binaries with the leading-order (1.5PN) spin-orbit couplings of one body spinning is chaotic. Levin (2003) claimed that the 2PN Lagrangian dynamics of two black hole binary system having one spinning body in harmonic coordinates is chaotic. However, the method of finding parametric solutions indicates the absence of chaos in the corresponding 2PN ADM Hamiltonian dynamics (Gopakumar and Konigsd 2005). These facts show that the two formulations have different dynamic behaviors, although they have been proven to be approximately equivalent (Levin 2006). Wu and Xie (2010) designed a set of conjugate spin variables to re-express the PN spin Hamiltonian part of the binary system. An explicit advantage for the construction of the globally symplectic PN Hamiltonian formulation lies in a good understanding of the relationship between the integrability and the least number of first integrals. The dynamic differences between the two 2PN approximately formulations of spinning compact binaries have been clarified in recent references (Wu et al. 2015; Wu and Huang 2015; Chen and Wu 2016), and the details can be found in these references.

The above mentioned disputes about chaos of PN spinning compact binaries indicate that the chaotic detection relies mainly on the used indicators and reliable numerical integrators under the specified dynamic parameters and initial conditions (Levin 2006; Wu and Xie 2007, 2008; Zhong and Wu 2010; Mei et al. 2013a). The conventional explicit Runge-Kutta (RK) numerical integrator has terms of secular change in energy errors and cannot preserve the geometric structure of Hamiltonian systems in the longtime integration; whereas, geometric integrator such as symplectic integrators (Wisdom and Holman 1991; Zhong et al. 2010; Ni and Wu 2014; Su et al. 2016; Mei et al. 2013b) can efficiently preserve energy and the geometrical structure of phase space of Hamiltonian system, usually been applied to simulate the evolution of PN compact binary systems. Recently, Liu et al. (2016) presented a fourth-order extended phase space explicit symplectic-like methods regarding symmetric compositions of two triple products of the usual second-order leapfrog. Luo used Yoshida's triple product combined with a midpoint permuted map to extend the reference's (Liu et al. 2016) work for inseparable Hamiltonian (Luo et al. 2017). An optimized fourth-order Forest-Ruth-like symplectic method is designed to solve inseparable Hamiltonian systems. (Wu and Wu 2018)

As one of the geometric integrators, manifold correction schemes can effectively compel the numerical solutions deviate from the original hypersurface to return to the original surface and eliminate the spurious numerical chaos (Nacozy 1971; Baumgarte 1972). The pioneering work of the manifold correction schemes was done by Nacozy in 1971. He first adopts Lagrange multipliers to force a calculated orbit back to the actual orbit according to a least-squares shortest path. Due to the convenience of Nacozy's method, there has been a lot of extension work (Chin 1995; Zhang 1996; Ascher 1997; Wu and He 2006; Han and Liao 2007; Ma and Long 2015). In particular, with the help of the integral invariant relation (Szebehely and Bettis 1971), this idea is extended to correct slowly-varying Keplerian energy, Laplace vector or orbital angular momentum integral for a perturbed Keplerian problem or each of multiple bodies in the planetary dynamics so as to drastically decrease the integration errors in various orbital elements (Fukushima 2003; Wu et al. 2007; Ma et al. 2008). Fukushima (2003) proposed the method of manifold correction with dual scaling for exact consistency of the Kepler energy, angular momentum vector and Laplace vector. The velocity scaling method for approximately preserving the Keplerian energy (Wu et al. 2007) is viewed as a direct extension of Nacozy's idea itself. Then, in the similar way, the velocity correction method to the Keplerian energy and Laplace integral is constructed (Ma et al. 2008). Moreover, Wang et al. (2016a; 2018) designed the velocity scaling method for restricted three-body problems; the numerical results indicated that the correction method is powerful for restraining the growth of integration errors.

As we have known, the conservative PN Hamiltonian formulation of spinning compact binaries has six motion integrals (including the total energy, three components of the total angular momentum and two constant lengths of spins), and we extend Nacozy's approach to the application of relativistic two-body problem (Zhong and Wu 2010), where the low-order RK5 method was taken as the basic integrator. The dual-scaling method of manifold correction was designed to numerically investigate the dynamics of spinning compact binaries. Numerical results show that it is very successful in repressing the linear growth of integration errors of RK5 method in long integration time, but the cost of computational time is expensive. When the two body spins are added in motion equation, the convergence rate during resolving the equations with respect to the scaling factors is much slower than the non-spinning binaries due to the influence of the complicated precession motion, with a vast number of iterations. Athough some optimization techniques are adopted in the program, it still needs much additional computational time. In a certain sense, our manifold correction method for PN Hamiltonian spinning compact binaries has high numerical precision and stability at the expense of the computation efficiency. We will focus on how to optimize manifold correction method with the help of modern computer technology to improve computational efficiency.

Graphics processing unit (GPU) has the capability of powerful floating-point operations and been industrialized rapidly in recent years, it is viewed as the most popular hardware accelerator in parallel computing. Since the Compute Unified Device Architecture (CUDA) was first introduced by NVIDIA as a parallel computing platform and software programming model, GPU has become more formidable and generalized with excellent performance and high efficiency, and has been widely used in flow simulations (Elsen et al. 2008; Rossinelli et al. 2010), molecular dynamics (Juba and Varshney 2008), electromagnetic wave, seismic wave propagation and other fields of research (Shams et al. 2010; Murray 2012; Zhang et al. 2015; Wang et al., 2016b). For example, Wang et al. (2016b) designed the GPU-accelerated finite-difference time-domain method for simulating EM wave propagation in the dielectric media (Wang et al. 2016b). Murray (2012) presented the GPU-accelerated explicit RK method for solving stiff differential equations, and so on. In addition, Moore et al. (2011) described a parallel hybrid symplectic integrator for simulating the outer solar system evolution based on the NVIDIA CUDA (Moore et al. 2011). These studies have confirmed that GPU can offer better application acceleration performance in settling massively parallel computations over CPU implementation. Thus, it is desirable for us to decrease the computational time by employing GPU accelerated manifold correction RK scheme, and maintain the numerical reliability, stability, and high precision in long term integration.

In this paper, one of our main objectives is to design a CUDA-implemented manifold correction RK scheme on GPU for modeling the long term evolution of the spinning compact black hole binaries, and adopt some optimization techniques to obtain a greater computational efficiency in comparison with the original RK method on CPU. Another objective is apprising how the orbital dynamical behaviour is affected by the spin-induced quadrupole-monopole (QM) interaction effect. The rest of this paper is organized as follows. In Section 2, PN Hamiltonian of spinning compact black hole binaries is introduced. Then, the manifold correction RK method is retrospected in brief. In Section 3, the GPU programming model and implementation of RK method based on CUDA is described, and some numerical comparisons (including speedup performance comparison for phase space scans) are also arranged to attest the feasibility and efficiency of parallelizing RK method in spinning compact binaries. The influences of the spin-induced QM for conservative black binaries on the orbital time and gravitational waves are evaluated in section 4. Finally, Section 5 summarizes our conclusions. Geometric units $c = G = 1$ are throughout the work.

## 2. Physical model and the manifold correction RK method

For a relativistic system of spinning black hole pairs, $m_1$ and $m_2$ ($m_1 \le m_2$) are the masses of two bodies, the total mass $M = m_1 + m_2$, the reduced mass $\mu = m_1 m_2 / M$, mass parameter $\eta = \mu / M$, and mass ratio $\beta = m_1 / m_2$. In the center of the mass frame, the relative position $\mathbf{r}$ and its magnitude $r = |\mathbf{r}|$ are measured in terms of $M$, and its canonical relative momentum $\mathbf{p}$ is measured in terms of $\mu$. The unit vector is $\mathbf{n} = \mathbf{r} / r$, and the spin vector of each body is $\mathbf{S}_i \left( i = 1, 2 \right)$. For an illustration, $\mathbf{r}$, $\mathbf{p}$ and $\mathbf{S}_i$ are all vectors in $\mathbb{R}^3$. The PN Hamiltonian formula of spinning compact binaries has the following form:

$$H(\mathbf{r}, \mathbf{p}, \mathbf{S}_1, \mathbf{S}_2) = H_N + H_{(1PN,2PN,3PN)} + H_{1.5PNSO} + H_{2PNSS} + H_{2.5PN}, \tag{1}$$

Where $H_N$, $H_{PN}$, $H_{SO}$, $H_{SS}$ and $H_{2.5PN}$ denote the pure orbital part, PN corrections, the spin-orbit coupling, spin-spin coupling term, and the gravitational radiated term, respectively. For simplicity, the detail expressions except spin-spin coupling term aren't listed here; they can be found in the references (Buonanno et al. 2006; Levin et al. 2011). The spin-spin part Hamiltonian reads:

$$H_{SS} = H_{S_1 S_1} + H_{S_1 S_2} + H_{S_2 S_2} \tag{2}$$

$$H_{S_1 S_2} = \frac{1}{\mathbf{r}^3} \left[ 3\ \mathbf{S}_1 \cdot \mathbf{n}\ \ \mathbf{S}_2 \cdot \mathbf{n}\ -\ \mathbf{S}_1 \cdot \mathbf{S}_2\ \right] \tag{3}$$

$$H_{S_1 S_1} = \frac{1}{2\mathbf{r}^3 \beta} \left[ 3\ \mathbf{S}_1 \cdot \mathbf{n}\ \ \mathbf{S}_1 \cdot \mathbf{n}\ -\ \mathbf{S}_1 \cdot \mathbf{S}_1\ \right] \tag{4}$$

$$H_{S_2 S_2} = \frac{\beta}{2\mathbf{r}^3} \left[ 3\ \mathbf{S}_2 \cdot \mathbf{n}\ \ \mathbf{S}_2 \cdot \mathbf{n}\ -\ \mathbf{S}_2 \cdot \mathbf{S}_2\ \right] \tag{5}$$

The second spin-spin coupling $H_{S_1 S_2}$ is valid for all bodies (Barker and O'Connell 1979) and the other two spin expressions ($H_{S_1 S_1}$ and $H_{S_2 S_2}$) denote the quadrupole-monopole terms, and are valid only for black hole binaries system (Damour 2001). They originate from the interaction of the monopole $m_2$ with the spin-induced quadrupole moment of the body $m_1$ and vice versa. The Hamiltonian canonical equations of motion for conjugate variables $(\mathbf{r}, \mathbf{p})$ can be written as:

$$\frac{d\mathbf{r}}{dt} = \frac{\partial H}{\partial \mathbf{p}} \qquad \frac{d\mathbf{p}}{dt} = -\frac{\partial H}{\partial \mathbf{r}} + \mathbf{a}_{2.5PN}, \tag{6}$$

where the relative acceleration $\mathbf{a}_{2.5PN}$ corresponds to the contribution of the acceleration from the $H_{2.5PN}$ radiation reaction in ADM coordinates. Levin et al. (2011) had transformed the 2.5PN radiation reaction expression of Lagrangian equation of motion into Hamiltonian coordinates read-to easy equation that governs the general black hole binaries.

Additionally, the time evolution of the spin variables is:

$$\frac{d\mathbf{S}_i}{dt} = \frac{\partial H}{\partial \mathbf{S}_i} \times \mathbf{S}_i = \Omega_i \times \mathbf{S}_i, \tag{7}$$

where

$$\Omega_1 = (2 + \frac{3}{2\beta})\frac{\mathbf{L}}{r^3} + \frac{1}{\eta r^3}\left[3\mathbf{n}(\mathbf{S}_2 \cdot \mathbf{n}) - \mathbf{S}_2\right] + \frac{3}{r^3}(1 + \frac{1}{\beta})^2 \mathbf{n}(\mathbf{S}_1 \cdot \mathbf{n}) \tag{8}$$

$$\Omega_2 = (2 + \frac{3\beta}{2})\frac{\mathbf{L}}{r^3} + \frac{1}{\eta r^3}\left[3\mathbf{n}(\mathbf{S}_1 \cdot \mathbf{n}) - \mathbf{S}_1\right] + \frac{3}{r^3}(1 + \beta)^2 \mathbf{n}(\mathbf{S}_2 \cdot \mathbf{n}). \tag{9}$$

For Hamiltonian system (1), there are six conserved quantities involving the total energy, three angular momentum components of the total angular momentum vector $\mathbf{J}$, and two constant unit lengths of spins $\hat{\mathbf{S}}_i \equiv 1$. We will now briefly review the dual-scaling manifold correction RK5 method for spinning compact binaries, namely a momentum-position scaling scheme for complete consistency of both the total energy and the magnitude of the total angular momentum. Assuming $\left[\mathbf{r}, \mathbf{p}, \hat{\mathbf{S}}_1, \hat{\mathbf{S}}_2\right]$ denotes a true solution of the system at time t, while $\left[\mathbf{r}^*, \mathbf{p}^*, \hat{\mathbf{S}}_1^*, \hat{\mathbf{S}}_2^*\right]$ and $\left[\bar{\mathbf{r}}, \bar{\mathbf{p}}, \hat{\bar{\mathbf{S}}}_1, \hat{\bar{\mathbf{S}}}_2\right]$ denote the computed solution and its more accurate solution at the same time t by using RK5 scheme to integrate the evolution equations, respectively. Due to the presence of various errors in the computational procedure, the computed solutions deviate from the true ones. In this sense, it is done via spatial scale transformations to the computed solution, in which scale factors are determined by constraining the solution $\left[\bar{\mathbf{r}}, \bar{\mathbf{p}}, \hat{\bar{\mathbf{S}}}_1, \hat{\bar{\mathbf{S}}}_2\right]$ on the proper integral surfaces. Thus, a more accurate solution, called the corrected solution, may become a good approximation to the true solution.

In the practical implementation, the two spin parameters are firstly corrected. In terms of the constant unit lengths of spins, a simple corrected scheme is adopted.

$$\hat{\mathbf{S}}_i \approx \hat{\bar{\mathbf{S}}}_i = \hat{\mathbf{S}}_i^* / S_i^* \tag{10}$$

At once, the constant unit lengths of spins are satisfied exactly at each integration step. Secondly, the momentum-position scaling scheme are adopted for corrections to both the computed total energy $H^*$ and the computed total angular momentum $\mathbf{J}^*$. The scale factor ($\alpha$, $\beta$) is derived from the least-squares correction of the total energy and the magnitude of the total angular momentum.

$$\mathbf{p} \approx \bar{\mathbf{p}} = \alpha \mathbf{p}^*, \quad \mathbf{r} \approx \bar{\mathbf{r}} = \beta \mathbf{r}^* \tag{11}$$

where the scale factor ($\alpha$ , $\beta$ ) is determined by the following minimizing error function

$$\Phi(\alpha,\beta) = w_1\left[H\left(\bar{\mathbf{r}},\bar{\mathbf{p}},\hat{\mathbf{S}}_1,\hat{\mathbf{S}}_2\right) - E_0\right]^2 + w_2\left\{\left[J\left(\bar{\mathbf{r}},\bar{\mathbf{p}},\hat{\mathbf{S}}_1,\hat{\mathbf{S}}_2\right) - J_0\right]^2\right\} \tag{12}$$

here $w_1$ and $w_2$ are two positive weight coefficients, $E_0$ and $J_0$ are the initial values of the total energy and the total angular momentum at the starting time, respectively. Equivalently, equation (12) vanishes

$$\begin{cases} \dfrac{\partial \Phi}{\partial \alpha} = 0 \\ \dfrac{\partial \Phi}{\partial \beta} = 0 \end{cases}. \tag{13}$$

The two scale factors are worked out by solving equation (13). Thus, the two integrals are always preserved rigorously at each integration step; the details can be seen in the references (Zhong et al., 2010).

## 3. The GPU-accelerated manifold correction RK method

### A. The CUDA-implemented manifold correction RK method

The scale factors can be obtained using Newton's iterative method for solving the above equations (13); however, when the spin effects are added in the motion equations, the iterative computational cost becomes very expensive so as to guarantee the numerical accuracy of the manifold correction RK5 method. In this paper, we design a parallel implementation scheme for solving equations (13). On the other hand, we also use GPU to accelerate the original RK5 methods for the motion equations of spinning compact binaries. The flowchart of GPU-based manifold correction RK5 method is illustrated in Fig. 1. The whole program procedure is carried out in three steps: Firstly, the computed solutions $\left[\mathbf{r}^*,\mathbf{p}^*,\hat{\mathbf{S}}_1^*,\hat{\mathbf{S}}_2^*\right]$ are obtained by using uncalibrated RK5 on GPU; Secondly, equation (13) is solved so as to obtain the scale factors ($\alpha$ , $\beta$ ) on GPU; Lastly, let the scale factor ($\alpha$ , $\beta$ ) act on the computed solutions $\left[\mathbf{r}^*,\mathbf{p}^*,\hat{\mathbf{S}}_1^*,\hat{\mathbf{S}}_2^*\right]$ on CPU. The host code is implemented to complete the initial conditions and dynamic parameters, device memory allocation and release, and data transfer between the host and the device. Shared memory is applied to optimize the performance of GPU-based correction RK method with CUDA. The CPU of the laptop is Intel Core i5 3230M and GPU NVIDIA GeForce GT 650M; the specifications are given in Table 1. The development environment is Microsoft Visual Studio 2013 (Community Edition) with CUDA toolkit 7.5 assembled, Windows 8 as the operating system.

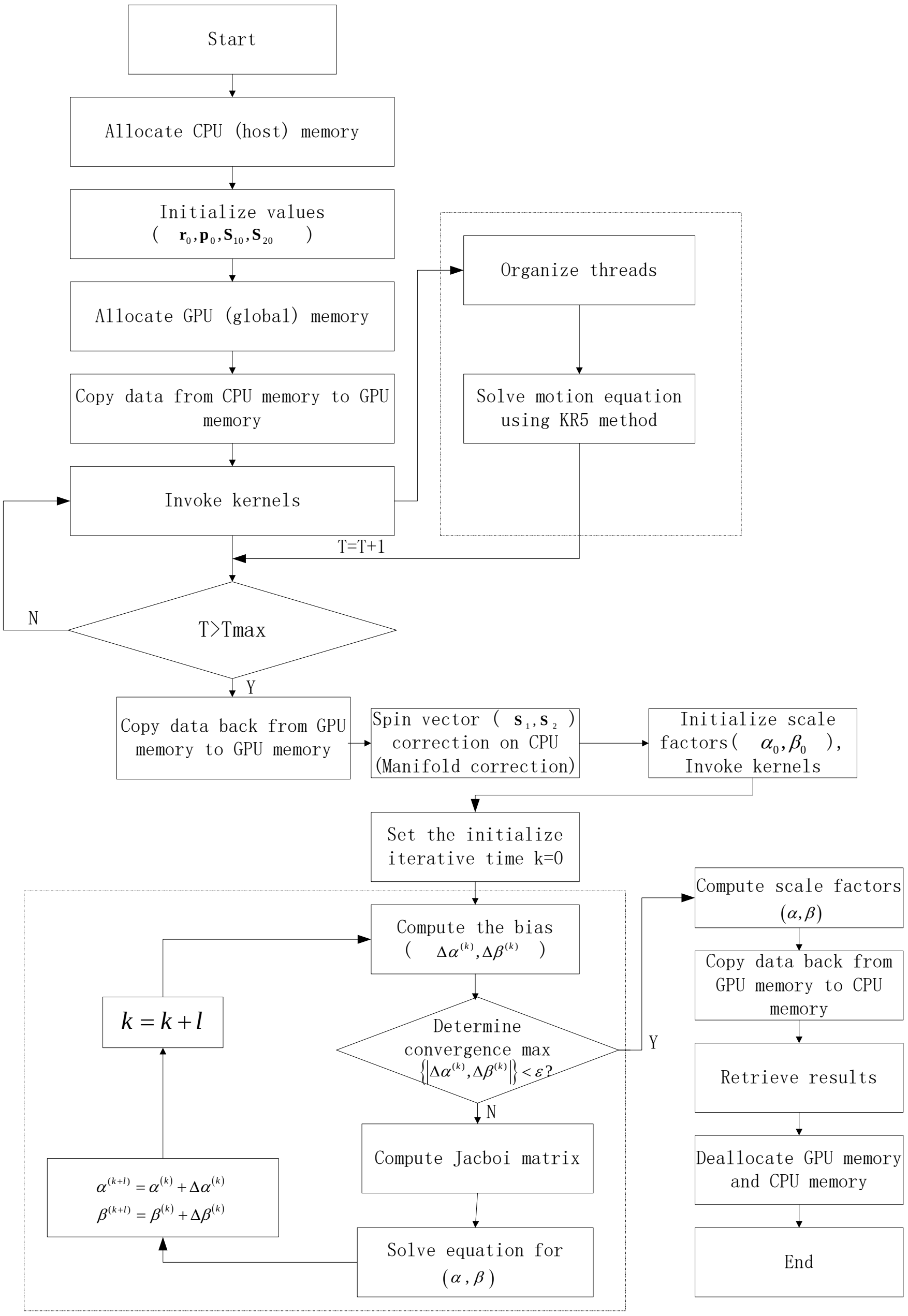

**Fig. 1** The flowchart of GPU-based manifold correction RK method.

**Table 1.** Specifications of NVIDIA 650M

| **Specification** | **GeForce GT 650M** |
|---|---|
| Chip | GK107 |
| CUDA cores | 384 |
| Processor clock | 835MHz |
| Memory clock | 900MHz |
| Memory size | 4096MB |

## B. Numerical experiments

In subsection B, the effect of the different iteration error limit of the scale factors in Newton's iterative method on the parallel speedup performance, and the numerical stability and accuracy of manifold correction RK5 method are tested. The integration time of each orbit is up to $1\times10^4$. Without loss of generality, various tested cases including the non-spinning binary and one with spins leading to chaos are considered. All computations are performed in a double precision environment.

When the spin effects and the 2.5PN radiation term are cut off in equation (1), the Hamiltonian system belongs to a typical perturbed Kepler problem in which the perturbation forces are from the PN corrections. The binary system is integrable and regular in the presence of four integrals, the energy and the angular momentum. The initial conditions and dynamic parameters for orbit $1\,(\beta,\mathbf{r},\mathbf{p})=\left(1/3, 20, 0, 0, 0, 0.2461, 0\right)$, the eccentricity e=0.1847. Orbit $2\,(\beta,\mathbf{r},\mathbf{p})=\left(1, 10.14, 0, 0, 0, 0.365, 0\right)$, corresponding to the eccentricity e=0.37. The eighth- and ninth-order Runge-Kutta-Fehlberg algorithm of variable step size (RKF8(9)) is taken as a higher precision reference integrator. Specifying the position error $\Delta x = x - \tilde{x}$, $x$ is reference solution given by RKF8(9) method, while $\tilde{x}$ is numerical solution. The energy errors and solution errors are shown in Fig. 2. For comparison, the RK5, an explicit and implicit mixed symplectic method S4A (Zhong et al. 2010; Mei et al. 2013a, 2013b) based on the fourth-order symplectic integrator of Forest & Ruth (1990), and another explicit and implicit mixed symplectic method S4B (Zhong et al. 2010; Mei et al. 2013a, 2013b) associated to the optimized Forest-Ruth-like fourth-order algorithm (Omelyan et al. 2002) are also shown in Fig.2. It can be seen that the accuracy of energy errors by the GPU-based manifold correction RK5 method and CPU-based correction RK5 are in agreement with RKF8(9) method, far superior to the fourth-order symplectic methods and uncorrected KR5 method, and almost arrives the machine precision order of $10^{-16}$. On the other hand, we can found that the correction method is better to control the growth of position error with time. In addition, Table 2 presents the energy errors $\Delta E$ and elapsed time for each method applied to the test orbits, and gives the corresponding speedup ratio at diverse given iteration error limits, respectively. The speedup ratio is calculated to evaluate the practical application speedup performance, which is defined as the ratio between the elapsed times of manifold correction RK5 computation on CPU and on GPU in the simulation. The GPU-based manifold correction RK5 method has a negligibly small decrease in the computational time than the manifold correction RK5 method on CPU. Perhaps due to the absence of the spin effects, the physics model is relatively simple and the PN correction terms have no significant effect on the dynamic evolution of spinning compact binaries. Also, it hardly requires additional computation time for solving scale factors Newton iteration method. For non-spinning compact binary system, these numerical methods have nearly the same computational efficiency.

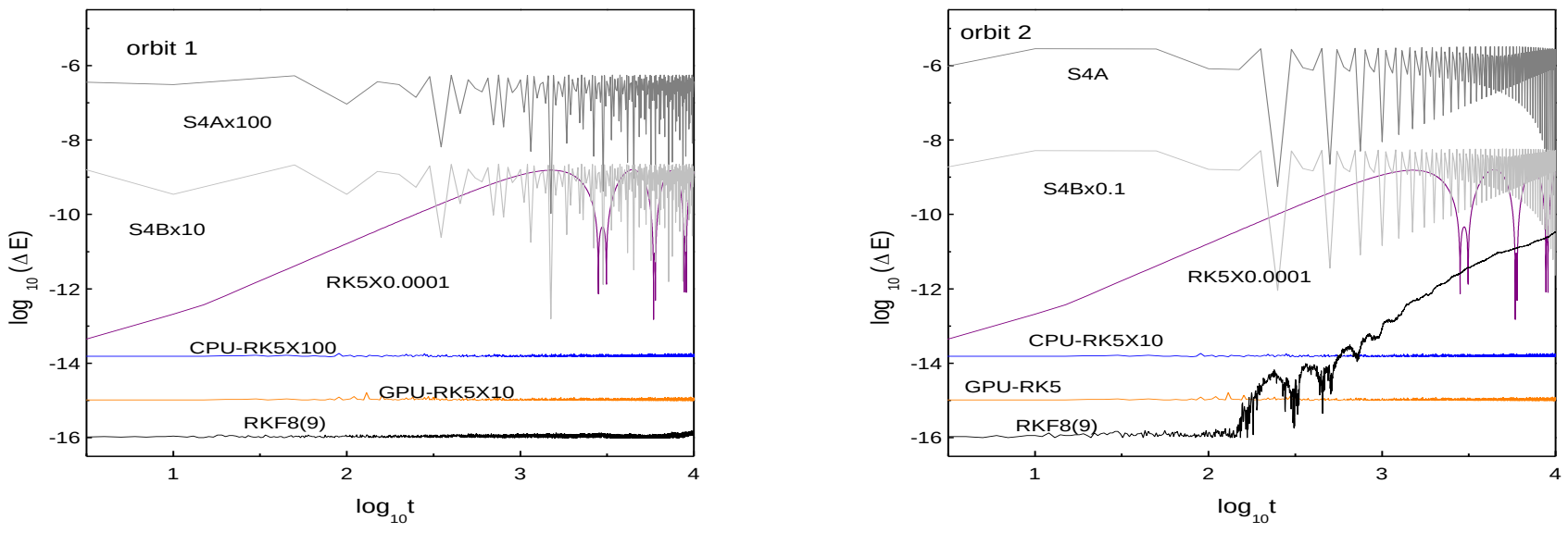

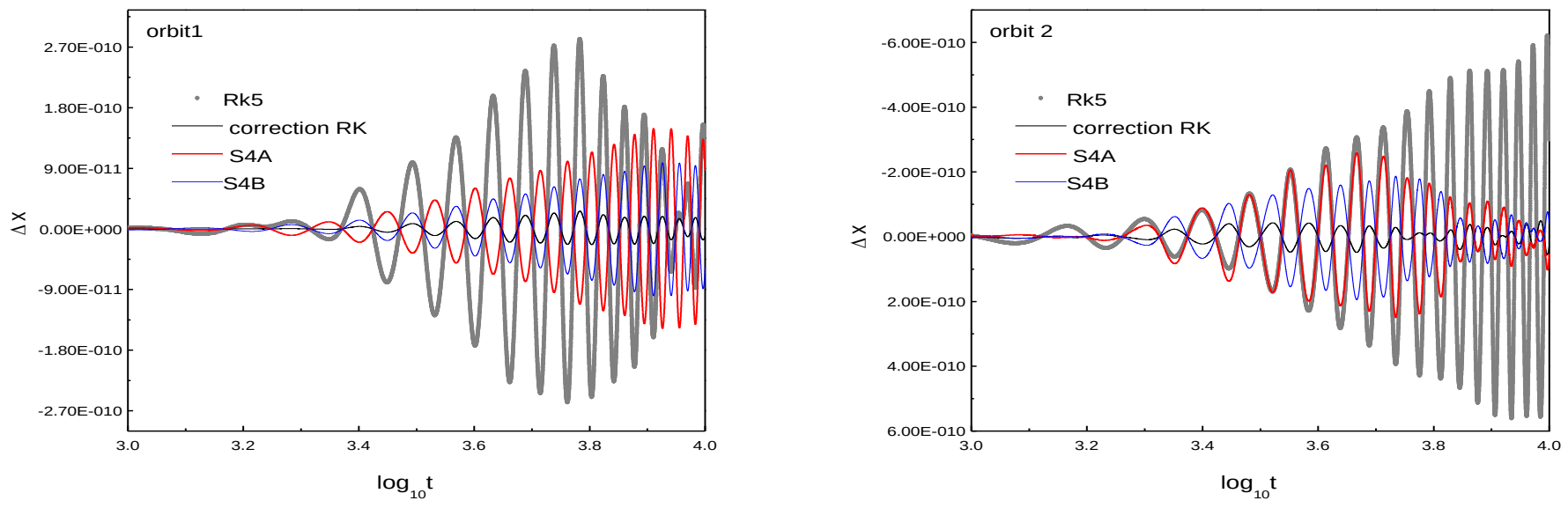

**Fig. 2** (color on line). The errors of the total energy and the solution errors for perturbed Kepler problem varying with time, the black line denotes the RKF8(9) method, the red line denotes the GPU-Rk5 method, the blue line is the CPU-Rk5 method, the purple is the RK5 method, the light gray is the optimized Forest-Ruth -like fourth-order explicit and implicit mixed symplectic algorithm, while the gray is the Forest-Ruth fourth-order explicit and implicit mixed symplectic method. Here the related symbol $RK5\times 0.0001$ denotes the plotted errors decreased by 10000 times compared with the real ones for RK5, while $CPU-RK5\times 10$ denotes the plotted error enlarged by 10 times compared with the real one for CPU-RK5. As illustrated, the symbol in Fig. 3 contains the similar meaning.

**Table 2.** Energy errors and computation times for various methods applied to test the orbits at iteration error limit $\varepsilon \le 10^{-10}$. Here case I denotes no spin binary and case II denotes the binary with two spin.

| **Cases** | **Orbit** | | **S4A** | **S4B** | **RK5** | **RKF8(9)** | **Correction RK5** | **Correction RK5 on GPU** | **Speed up** |
|---|---|---|---|---|---|---|---|---|---|
| Case I | 1 | $\Delta E$ | $10^{-8}$ | $10^{-10}$ | $10^{-7}$ | $10^{-15}$ | $10^{-15}$ | $10^{-15}$ | 1 |
| | | Time(s) | 4 | 4 | 4 | 6 | 4 | 4 | |
| | 2 | $\Delta E$ | $10^{-6}$ | $10^{-7}$ | $10^{-6}$ | $10^{-12}$ | $10^{-15}$ | $10^{-15}$ | 1 |
| | | Time(s) | 4 | 5 | 4 | 6 | 4 | 4 | |
| Case II | 1 | $\Delta E$ | $10^{-3}$ | $10^{-4}$ | Linear growth | $10^{-5}$ | $10^{-15}$ | $10^{-15}$ | 6.85 |
| | | Time(s) | 9 | 13 | × | 20 | 89 | 13 | |
| | 2 | $\Delta E$ | $10^{-3}$ | $10^{-4}$ | Linear growth | $10^{-4}$ | $10^{-11}$ | $10^{-11}$ | 7.5 |
| | | Time(s) | 9 | 14 | × | 18 | 45 | 6 | |

When considering the spin effect and pruning the 2.5PN radiated term, the system remains conservative and integrable. The spin parameters and initial configurations for orbit 1: $\mathbf{S}_1 = (\sin 60^\circ, 0, \cos 60^\circ)$, $\mathbf{S}_2 = (\sin 45^\circ, 0, \cos 45^\circ)$, the spin parameters $\chi_1 = \chi_2 = 0.75$; orbit 2 $\mathbf{S}_1 = (0.13036, 0.262852, -0.955989)$, $\mathbf{S}_2 = (0.118966, -0.13459, -0.983734)$, $\chi_1 = \chi_2 = 1$. Once the spin effects are added in motion equations, the dynamics of the binary system becomes very complicated and the orbit planes occur precession motion along the z direction, as shown in Fig. 3(a) and 3(c). The fast Lyapunov indicator (Wu et al. 2006) with the power-law divergence of initially close trajectories in Fig. 3(b) proves that orbit 1 is regular, while the exponential-law divergence of FLI in Fig. 3(d) shows that orbit 2 is chaotic.

The computational times and the energy errors are also presented in Table 2. The numerical results show that the spin effects have a significant effect on the quality of the symplectic integrators and the RKF8(9) method. Their accuracy only arrives $\times 10^{-4}$ order of magnitude for spinning compact binaries, and is far inferior to that of the RK correction method. Fig. 4 presents the energy errors of RK method on GPU varying with time at different iteration error limit of the scale factors, and the CPU is also given at iteration error limit $\varepsilon \le 10^{-10}$. We can see that the iteration convergence of the manifold correction method depends on the selection of error limit. If the iterative error limit $\varepsilon \le 10^{-6}$ or $\varepsilon \le 10^{-8}$ program code is implemented, then the energy error gradually increases linearly with the integration time increasing as shown in the blue and red line in Fig. 4, especially for the chaotic orbit, that is, the manifold correction RK methods can't retain the long-term numerical stability. Lots of numerical experiments show that it is desirable when the iterative error limit $\varepsilon \le 10^{-10}$ is shown by the black line in Fig. 4. Hence, the iterative error limit $\varepsilon$ is set $\le 10^{-10}$ in the following numerical simulations. The accuracy of position error is decreased, but the correction method is still best compared with other methods, as shown in Fig. 4. The speedup achieves around 7 at the iterative error limit $\varepsilon \le 10^{-10}$ as shown in Table 2, and we all know that there are many factors that can negatively affect the accelerating effect for spinning compact binaries, for example, the spin effect, orbital classification, the GPU device properties and the developer's programming levels otherness, and so on. The test results indicate that the proposed GPU-based manifold correction RK5 method implemented with CUDA can obtain a decent acceleration effect with sufficient accuracy when compared to only CPU-based manifold correction RK5 method.

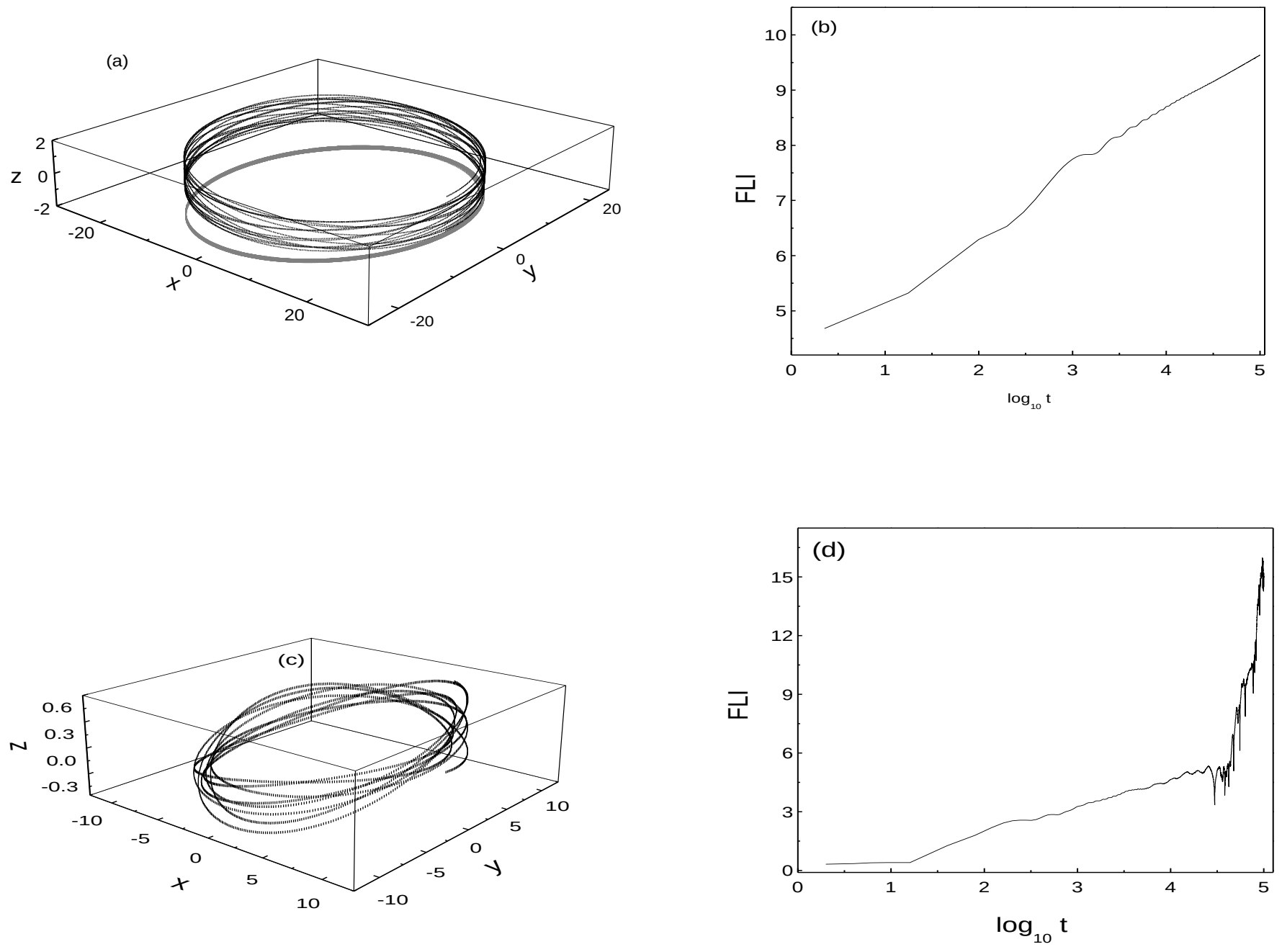

**Fig. 3** Panel (a) is a three-dimensional view of orbit 1, and panel (b) relates to it's corresponding FLIs. Panels (c) and (d) are similar to (a) and (b), but for orbit 2, respectively.

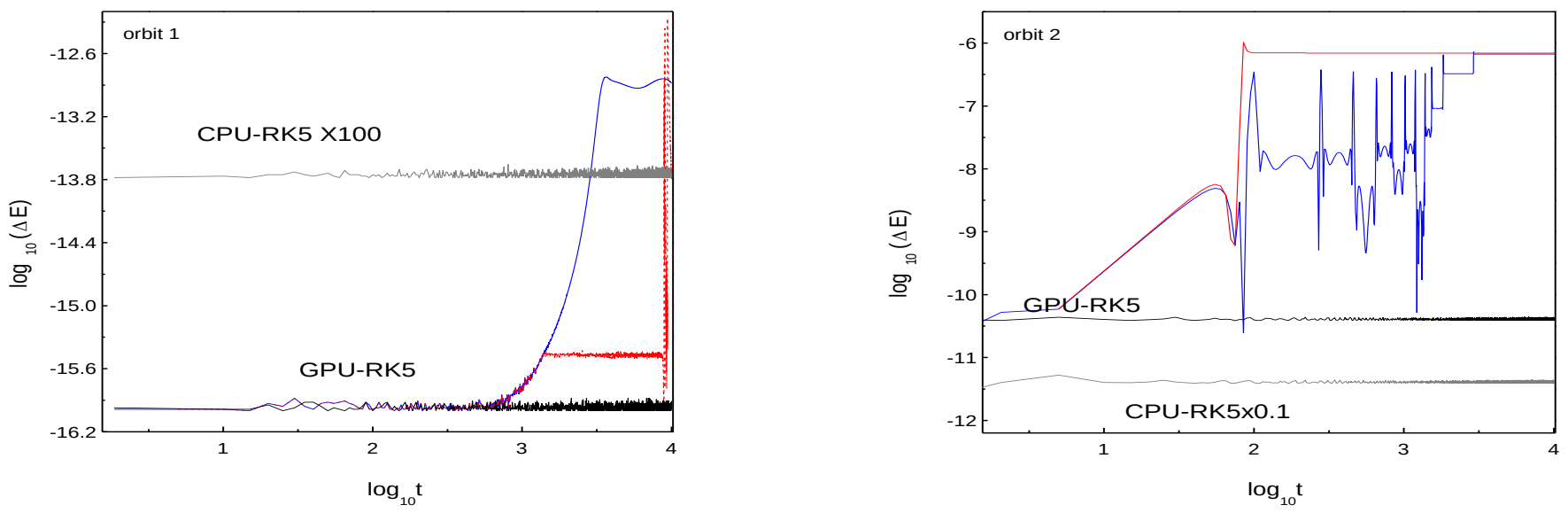

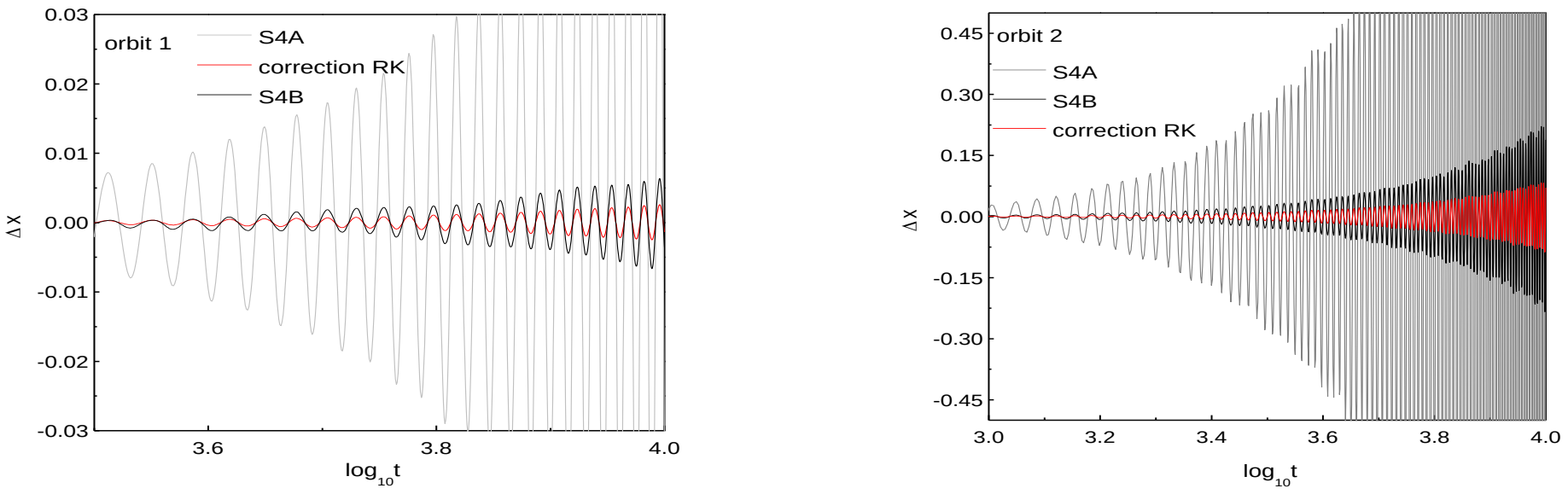

**Fig. 4** (color on line) Similar to Fig. 2 but for conservative spinning compact binaries, the gray line denotes the CPU-based manifold correction RK5 method, while black, red and blue lines indicate the GPU-accelerated manifold correction RK5 method under the iteration error limit of the scale factors $\varepsilon \le 10^{-10}$ , $\varepsilon \le 10^{-8}$ and $\varepsilon \le 10^{-6}$ , respectively.

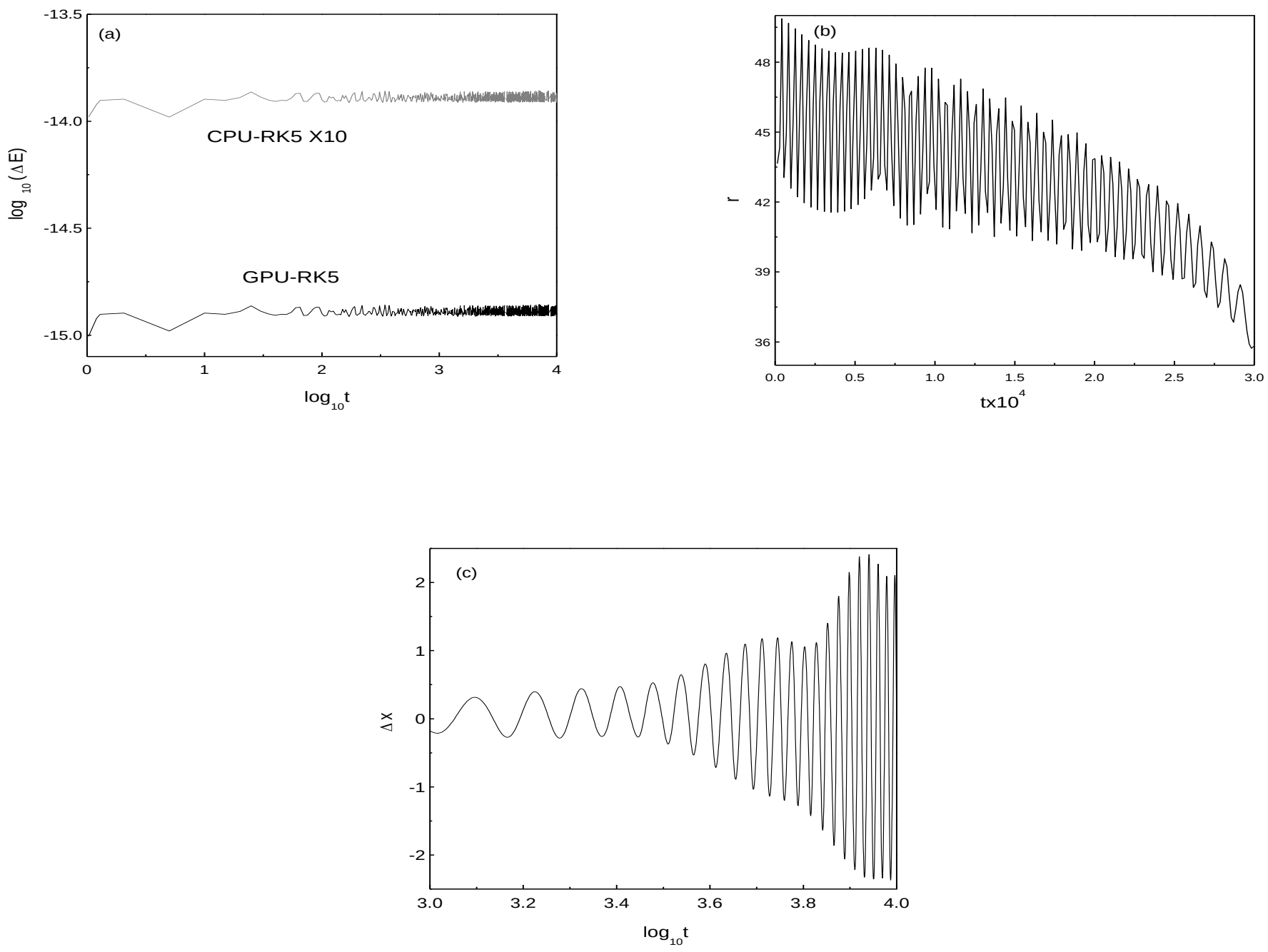

**Fig. 5** (a) Similar to Fig. 4 but for non-conservative spinning compact binary at the iterative error limit of the scale factors $\varepsilon \le 10^{-10}$ , (b) refers to the magnitude of relative position $\mathbf{r}$ decreasing with time, (c) plots the errors between the solutions of RKF8(9) and corrected RK5, the errors between the solutions of RKF8(9) and uncorrected RK5 are not plotted because uncorrected RK5 fails to work during such a long integration time.

We further test the effectiveness of the correction scheme for non-conservative binary system. When the 2.5PN gravitational radiated term is included in equation (1), the Hamiltonian system becomes non-conservative and non-integrable, and the energy of the system changes with time. For a dissipative system, we can use the Hamiltonian from the integral invariant relation as the accurate reference value to test the accuracy of numerical integration; see the reference (Luo and Wu 2017) for more details. The initial conditions and dynamic parameters of orbit 3: $(\beta, \mathbf{r}, \mathbf{p}) = (1, 50, 0, 0, 0, 0.140, 0)$ , $\mathbf{S}_1 = (\sin 45^\circ, 0, \cos 45^\circ)$ , $\mathbf{S}_2 = (\sin 30^\circ, 0, \cos 30^\circ)$ , $\chi_1 = 1.0$ , $\chi_2 = 0.85$ . Although the instantaneous energy of the non-conservative spinning compact binary continuously changes with time, the relative energy errors have no secular linear growth as shown in Fig. 5(a), as the correction method in conservative case does. Fig. 5(b) describes the magnitude of the relative position $\mathbf{r}$ of binary damping with time, and the computation cannot continue when the $r$ value decays to a critical value. Fig. 5(c) plots the errors between the solutions of RKF8(9) and corrected RK5, the errors between the solutions of RKF8(9) and uncorrected RK5 are not plotted because uncorrected RK5 fails to work during such a long integration time. The computation efficiency and speedup ratio for non-conservative case are almost in accordance with the conservative case.

### C. Speedup performance comparison for phase space scans

In this subsection, we continue to evaluate the speedup performance of the GPU-based correction RK method by phase space scanning for chaos and order with fixed some initial parameters and FLIs. The presence of chaos is decided by a multiple combination of dynamics parameters and initial conditions of two objects. In the first place, we trace a transition to chaos as one of the two spin angles is varied initially, and the dynamic parameters and initial conditions are fixed: $(\beta, \mathbf{r}, \mathbf{p}) = \left(1/3, 5.65, 0, 0, 0, 0.765, 0\right)$, spin angle $\theta_1 = \frac{\pi}{2}$, let spin angle $\theta_2$ run from 0 to $\pi$ with a span of intervals $\Delta\theta_2 = 0.01$, hence 314 orbits are integrated, and the value of FLI for each orbit is obtained when the integration time is up to $10^5$. Numerous numerical simulations indicate 5.8 as a critical value of FLIs between order and chaos. We can deduce from Fig. 6(a) that the presence of chaos is located mainly at large spin angle $\theta_2$ range. If the initial spin angle $\theta_1 = \frac{\pi}{2}$ is replaced only with $\theta_1 = 0^\circ$, chaos will not occur, as shown in Fig. 6 (b). Fig. 7 shows a precise phase space structure of order and chaos on the $(\theta_1, \theta_2)$ plane, the numbers of integrated orbit is $314 \times 314$, the unstable unbounded or merged orbits are get rid of in simulation, the unbounded unstable dynamic of orbits is discussed at length in the reference (Huang and Wu 2014). The total implementation times of correction RK method on CPU and GPU are listed in Table 3. Compared to the time implemented in CPU, the correction RK method on GPU can greatly save the computing time and improve computational efficiency, and the speed up is more than 10.

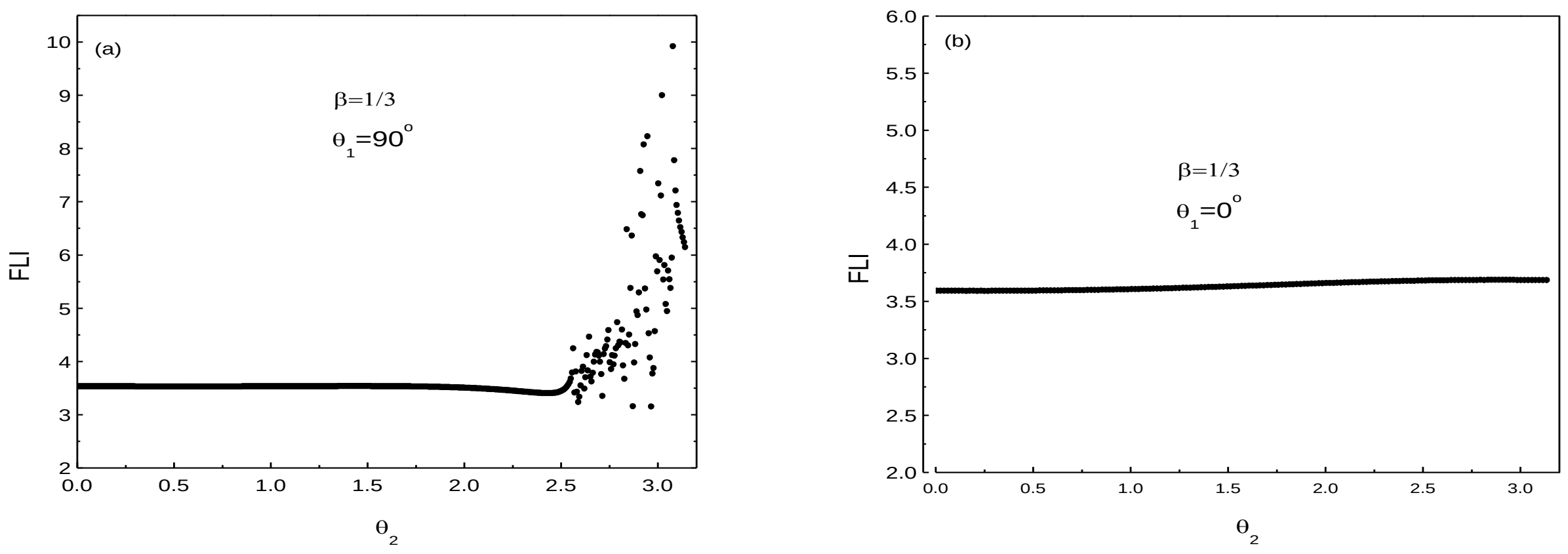

**Fig. 6** FLI as a function of initial spin angle $\theta_2$ with fixed initial states $(r = 5.65, p_y = 0.765)$ and spin parameters $\chi_1 = \chi_2 = 1$.

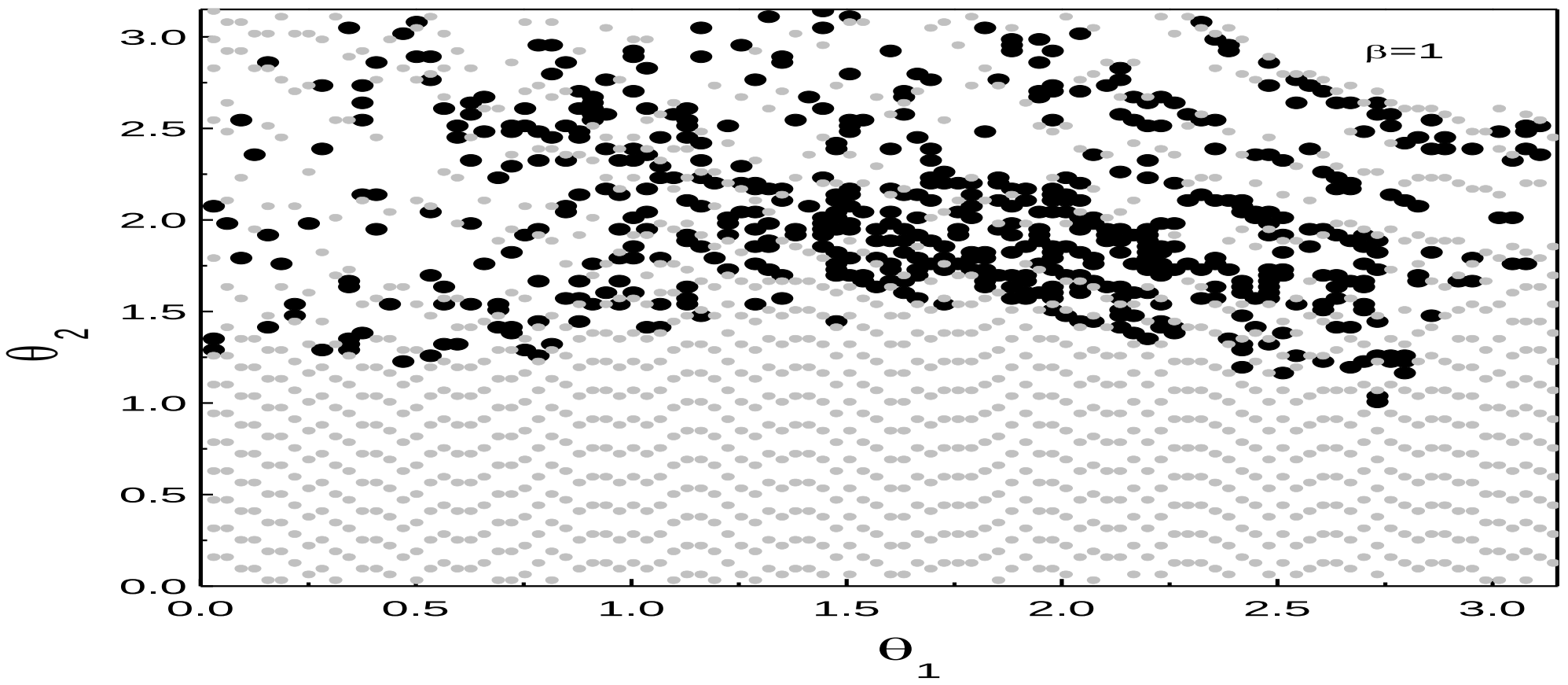

**Fig. 7** Similar to Fig. 6, but scans of a group of initial points on the ( $\theta_1$ , $\theta_2$ ) plane for chaos with fixed initial states $(r=10.14, p_y=0.365)$ and spin parameters $\chi_1=\chi_2=1$. Black areas with FLIs $>5.8$ indicate chaos, and gray areas with FLIs $<5.8$ show order.

**Table 3.** Computation time and Speedup ratio for phase space scans on the ( $\theta_1$ , $\theta_2$ ) plane.

| Orbit numbers | CPU Times(s) | GPU Times(s) | Speedup |
|---|---|---|---|
| 314 | 1020 | 75 | 13.6 |
| 314*314 | 25000 | 2213 | 11.29 |

## 4. The dynamic evolution of black hole binary system and gravitational waves

In this section, we first evaluate how the dynamic behavior is modulated by the spin-induced quadrupole-monopole contributions for a conservative regular orbit 4. Its initial conditions and dynamic parameters $(\beta,\mathbf{r},\mathbf{p})=(1,10,0,0,0,0.274,0)$, the starting unit spin configurations $\mathbf{S}_1=(\sin\theta_1,0,\cos\theta_1)$, $\mathbf{S}_2=(\sin\theta_2,0,\cos\theta_2)$, the spin parameters $\chi_1=\chi_2=1$, the initial spin angle $\theta_1=1.0rad$, $\theta_2=1.325rad$. Fig. 8(a) plots a three-dimensional view of orbit 4, and the FLI with the power-law divergence of initially close trajectories in Fig. 8(b) proves that the orbit is regular. Fig. 8(c)-(f) describes the influences originated from the QM interaction contribution on the orbital dynamic evolution. In the absence of gravitational radiation dissipation, the orbital time evolution seems to become more disordered and looks like a nest when the QM interaction contributions are added in the motion equations as shown in Fig. 8(c); meanwhile, the geometry structure of phase space of Hamiltonian system also occurs in obvious distinction due to the QM interaction compared to the case of uncoupling QM contribution, as shown in Fig. 8(d). On the other hand, we can see from Fig. 8(e) and (f) that there are some important differences in that the orbital plane xoy tilts back and forth as it precesses about total angular momentum **J;** although the two bodies initiate the precess at different rates, they gradually approach overlapping each other with the QM interaction joining and the time increasing, as well as the inclination of the orbital plane with respect to the total angular momentum initial value $J_0$ augments as the distance of the two bodies decreases as shown in Fig. 8(f). Fig. 9 depicts the three spin components $s_x$, $s_y$ and $s_z$ of two bodies time evolution in “spin space” in a Cartesian coordinate system. It can be found that there are still qualitative varieties in the spin behavior between the cases of the QM interaction contributions being turned off and turned on.

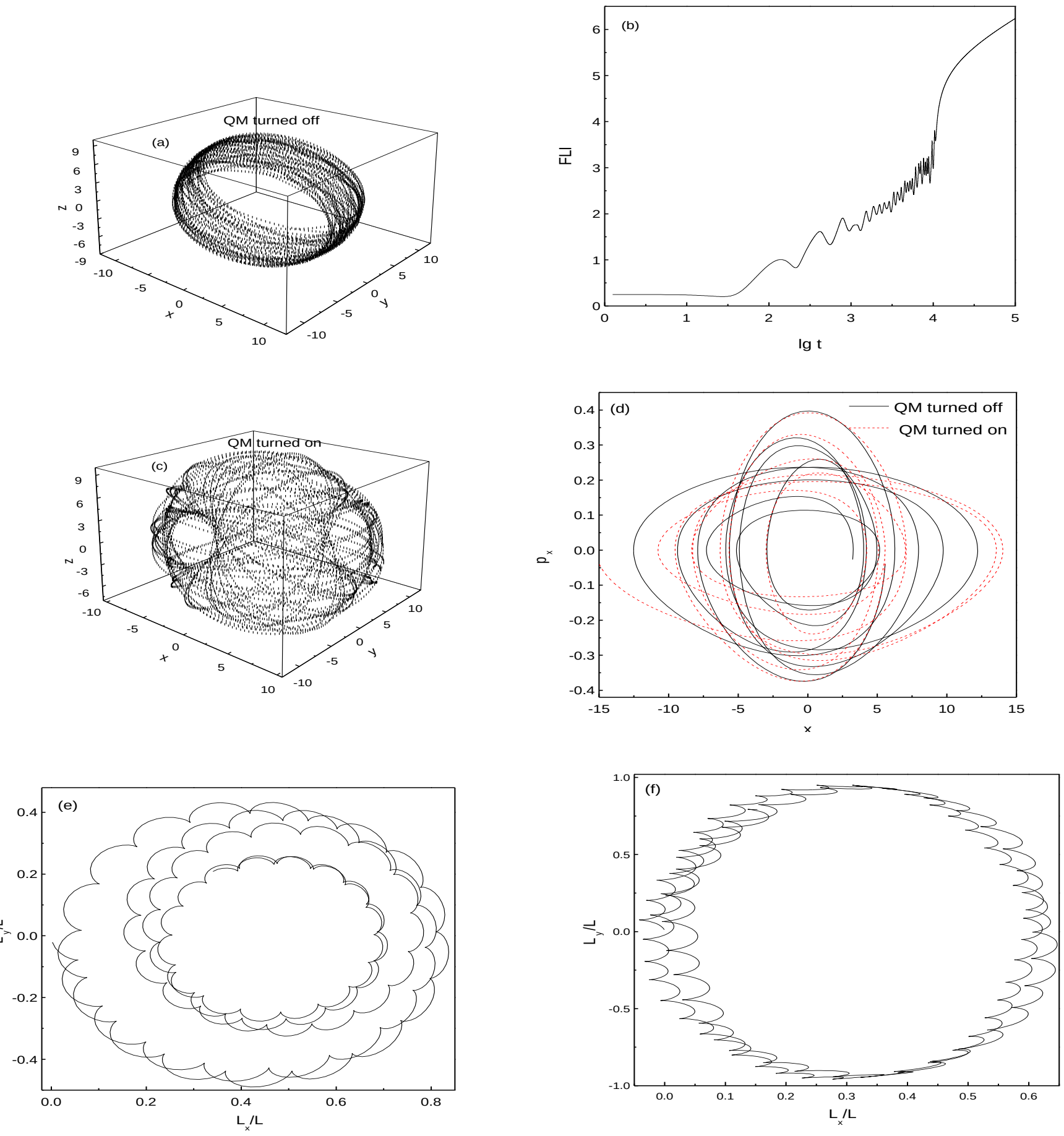

**Fig. 8** (color online) (a) and (c) are the three-dimensional view of orbit 4 with mass ratio $\beta = 1$, (b) relates to it's corresponding FLIs. (d) is the projection of the orbit onto the $x - p_x$ plane, and the bottom ones relate to the complicated precession of the orbital angular momentum **L.** The left panels belong to the cases where the QM interaction contributions aren't considered, the right is contrary.

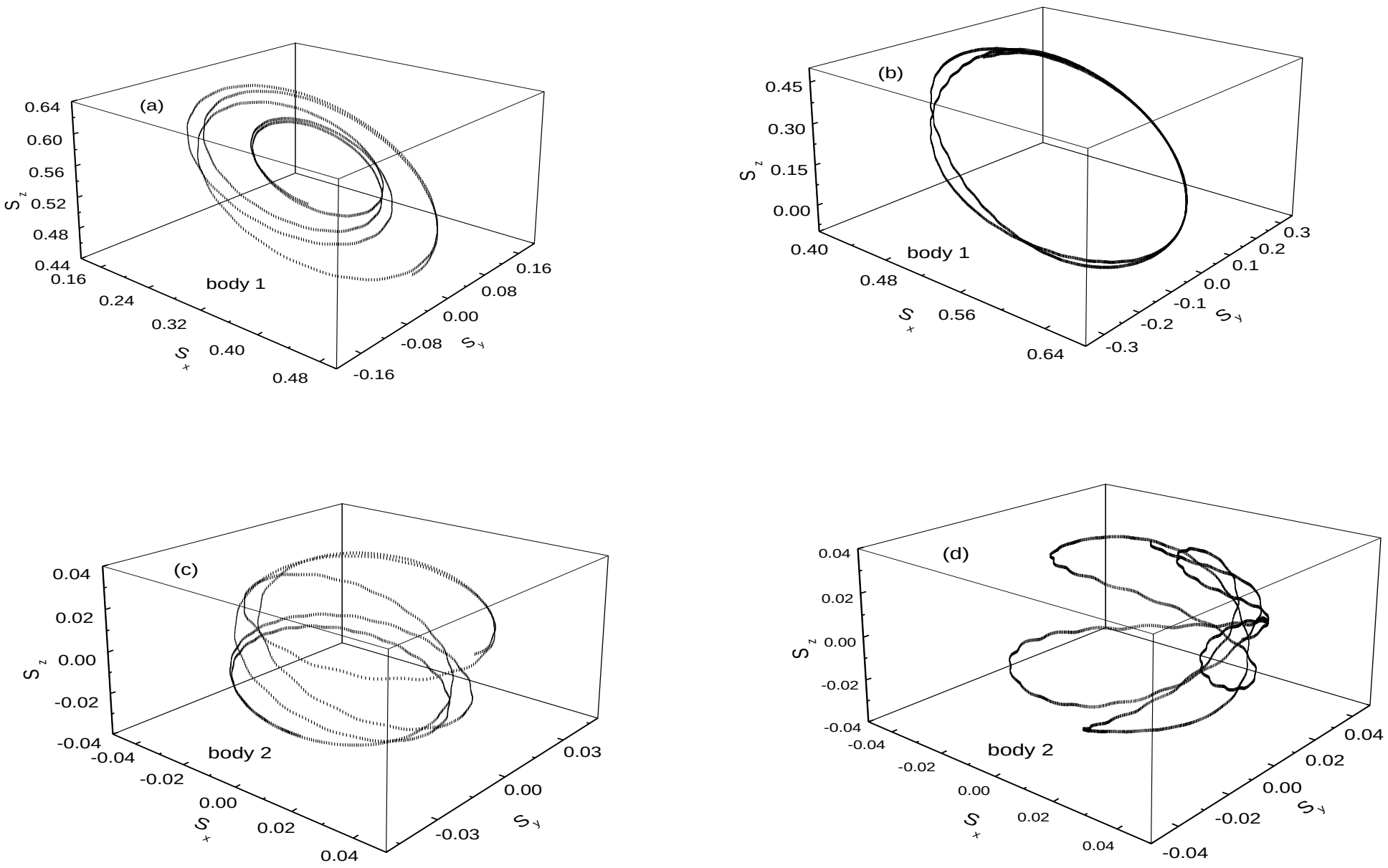

**Fig. 9** "Spin space" in Cartesian coordinate system showing the two bodies time evolution of $S_x$, $S_y$ and $S_z$, here the left denotes the case without considering the QM interaction contributions, the right panels is contrary.

The above results indicate that the spin-induced QM perturbation interaction exerts a non-negligible physical influence on the orbital dynamic evolution for the given system, and the system may be unstable and chaos might occur due to spin effect. Of course, these physical effects not only rely solely on QM perturbation interaction, but also on a complicated combination of all parameters and initial conditions. In future work, we will continue to investigate how the system transforms order into chaos owing to spin effect under some fixed conditions.

Next, we numerically evaluate how system dynamics characteristics exert different influences on the gravitational waveform. When the 2.5PN radiation interactions are considered, the conservative binary system becomes the non-conservative system. The two independent gravitational wave polarization states $h_+$ and $h_\times$ are given, with respect to the radiation frame $\left(\hat{\mathbf{N}},\hat{\mathbf{p}},\hat{\mathbf{q}}\right)$, as

$$
\begin{aligned}
h_+ &= \frac{1}{2}\left(\tilde{p}_i\tilde{p}_j - \tilde{q}_i\tilde{q}_j\right)h_{ij}^{TT} \\
h_\times &= \frac{1}{2}\left(\tilde{p}_i\tilde{q}_j + \tilde{q}_i\tilde{p}_j\right)h_{ij}^{TT}
\end{aligned}
\qquad (14)
$$

where $\hat{\mathbf{N}}$ is a triad of unit vectors pointing from the center of mass of the source to the observer, $\hat{\mathbf{p}}$ lies along the line of nodes, defined by $\hat{\mathbf{q}} = \hat{\mathbf{N}}\times\hat{\mathbf{p}}$, $h_{ij}^{TT}$ is the transverse-traceless (TT) part of the radiation field representing the deviation of the metric from the flat spacetime (Gopakumar and Iyer 2002). Fig. 10 presents the time dependence of the $h_+$ gravitational waves from the mass quadrupole moment of orbit 4, and for comparison, the gravitational waves from chaotic orbit 2 above mentioned in section 3 is also tested, where the triad of unit vectors are taken as : $\hat{\mathbf{p}} = (1,0,0)$, $\hat{\mathbf{q}} = (0,1,0)$ and $\hat{\mathbf{N}} = (0,0,1)$, respectively. There are two differences of gravitational waveform between the conservative and non-conservative cases; one is that the amplitude of the gravitational wave from non-conservative binary system is larger than that of conservative case, as shown in Fig. 10; the other is that the duration time of the gravitational radiation in non-conservative cases is very short as shown in Fig. 10(c) and (d). Moreover, as shown in Fig. 10(c) and (d), the gravitational waveform from regular orbit changes periodically and the amplitude is less by one order of magnitude than the chaotic case.

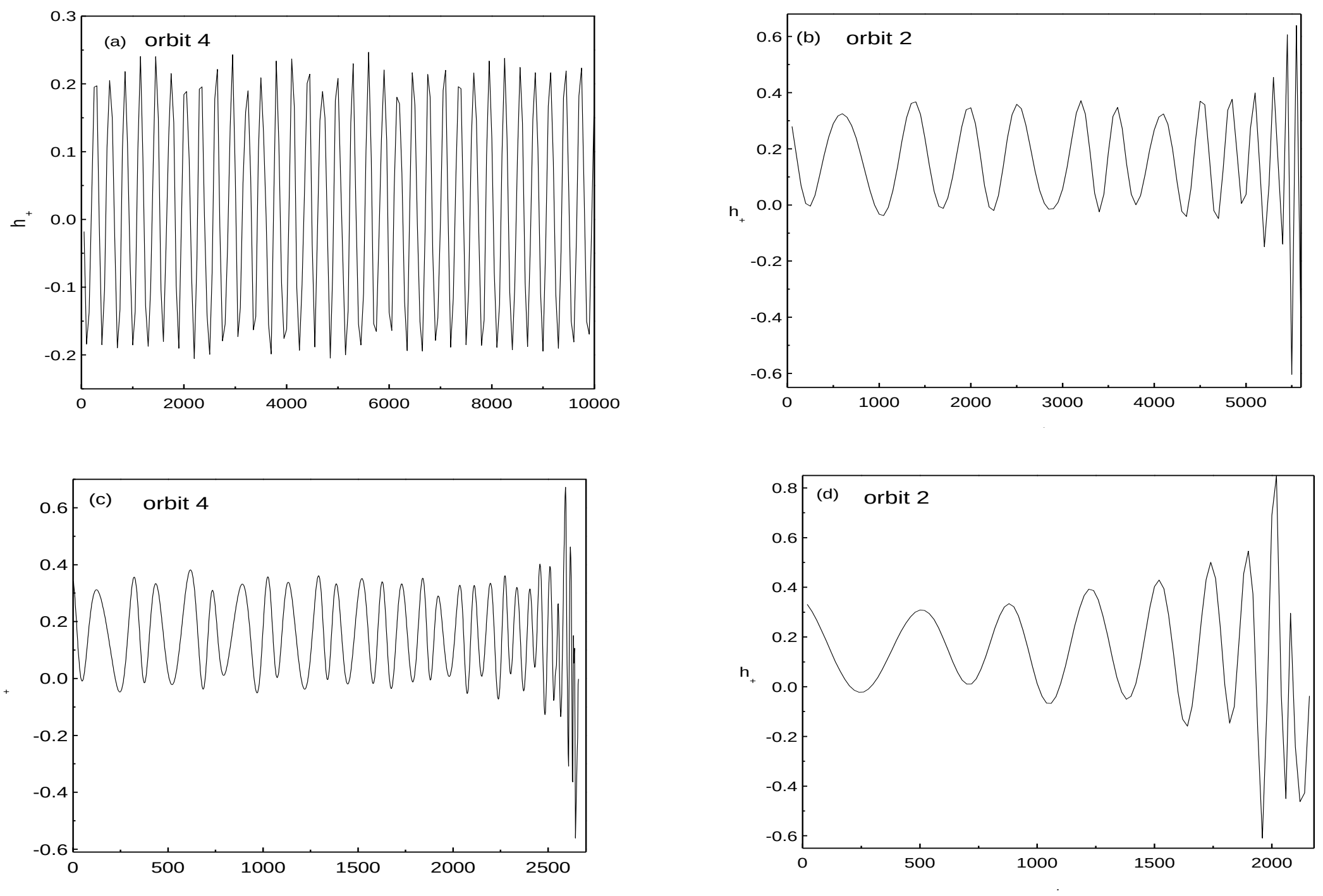

**Fig. 10** shows the comparisons of the influences of orbital classification on gravitational waveform. The top curves denote the gravitational radiation from the conservative binary system, and the bottom curves indicate the gravitational wave emit from non-conservative binary system.

# 5. Conclusion

We present a GPU acceleration of the manifold correction method for simulating the dynamic evolution of the PN Hamiltonian formulation of spinning compact black hole binaries. The numerical experiments confirm the feasibility of parallel computation on GPU for modeling time evolution of spinning binaries. The accuracy of GPU-based manifold correction method is consistent with the codes execution on CPU and the computational efficiency is satisfactory. It solves the bottleneck problem of low efficiency in long term integration for relative two body problem.

We observe the characteristics of time evolution of PN black hole binaries when the spin-induced quadrupole moment interactions are neglected and considered. The results show that the precession of the orbital angular momentum **L** gradually cancel each other and the spin behavior of every body has great differences compared to the case without coupling of quadrupole moment interactions.

**Acknowledgements** This research is supported by the Natural Science Foundation of China under Contract (Grant No.11563006,11165011) .